\documentclass[pra,twocolumn,amsmath,amssymb,floatfix,reprint]{revtex4-2}

\usepackage{xcolor}
\usepackage{graphicx}
\usepackage[normalem]{ulem}
\usepackage{dcolumn}
\usepackage{booktabs}
\usepackage{bm}

\begin{document}

\preprint{APS/123-QED}

\title{Finite-time thermal refrigerator in interacting Bose–Einstein Condensates}
\author{Joaquín I. Ganly$^{1}$}\email{ganlyjoaquin@gmail.com}
\author{Julián Amette Estrada$^{1,2}$}\email{julianamette@df.uba.ar}
  \author{Franco Mayo$^{1,3}$}\email{fmayo@df.uba.ar}
  \author{Augusto J. Roncaglia$^{1,3}$}\email{augusto@df.uba.ar}
  \author{Pablo D. Mininni$^{1,2}$}\email{mininni@df.uba.ar}
  
\affiliation{$^1$Universidad de Buenos Aires, Facultad de Ciencias Exactas y Naturales, Departamento de Física, Ciudad Universitaria, 1428 Buenos Aires, Argentina,}
\affiliation{$^2$CONICET - Universidad de Buenos Aires, Instituto de F\'{\i}sica Interdisciplinaria y Aplicada (INFINA), Ciudad Universitaria, 1428 Buenos Aires, Argentina.}
\affiliation{$^3$CONICET - Universidad de Buenos Aires, Instituto de F\'{\i}sica de Buenos Aires (IFIBA), Ciudad Universitaria, 1428 Buenos Aires, Argentina.}

\date{\today}

\begin{abstract}
We study a finite-time thermodynamic refrigeration cycle realized numerically in three-dimensional, weakly interacting Bose–Einstein condensates (BECs). The setup consists of three spatially separated condensates---system, piston, and reservoir---coupled through time-dependent potential barriers that implement compression, expansion, and contact strokes. Finite-temperature initial states are generated with the Stochastic Ginzburg–Landau equation, and the subsequent dynamics are evolved using the truncated Gross–Pitaevskii equation. To measure temperatures we use a momentum-space thermometry method that provides estimates for each condensate. We find that despite mass transfer and sound excitations, the protocol achieves successful cooling during consecutive cycles: the first cycle lowers its temperature by ~$20\%$, and a second cycle yields additional, though reduced, cooling, reaching a final ~$27\%$ cooling from the initial state. 
Our results show that interacting BECs can sustain finite-time quantum thermal cycles under realistic conditions, and provide a platform for exploring different refrigeration schemes, optimized control protocols, and shortcuts to adiabaticity.
\end{abstract}

\maketitle


\section{Introduction}
\label{sec:intro}

The study of ultracold quantum gases has become a central avenue for exploring nonequilibrium quantum phenomena at mesoscopic and macroscopic scales. Among these systems, atomic Bose–Einstein condensates (BECs) stand out due to their high coherence, tunable interactions, and the remarkable degree of dynamical control achieved in modern experiments \cite{Anderson, Myers, simmons2023thermodynamic}. These properties make BECs promising candidates for realizing quantum thermodynamic machines, in which fundamental notions such as work, heat, and efficiency are explored beyond the paradigmatic few-level setting.
The field of quantum thermodynamic machines \cite{Vinja, Binder2018, Bhattacharjee2021, Jordan2020, Mitchison, Niedenzu} seeks to understand how classical thermodynamic concepts translate to these kinds of systems. While substantial progress has been made for discrete few-level engines, extending these ideas to continuous and coherent many-body systems remains comparatively underdeveloped.

Recent studies have shown that time-dependent manipulation of trapping potentials can emulate the strokes of thermal machines, enabling controlled energy exchange between spatially separated regions of a condensate. In particular, Gluza et al.~\cite{Gluza} recently introduced a one-dimensional, non-interacting cooling cycle in which a dynamically driven ``piston'' alternately couples to a thermal reservoir and to a ``system'' region which is the target of the cooling process. However, the absence of interactions, the reduced dimensionality could restrict the applicability of these models to more realistic BEC platforms  \cite{Ruprecht, Fialko2012, Cangemi2023}.

Recent experimental and numerical work attempted to address some of these challenges. The equation of state of trapped condensates was examined in \cite{reyes2023carnot}, and an Otto cycle was realized experimentally in \cite{simmons2023thermodynamic, Rossnagel2016, Abah2012, Lindenfels2019, Koch2023}. Interacting BECs have also been incorporated into theoretical studies \cite{Dann2020, Camati2019, Hewgill2018, Elouard2017, Elouard2018, Nautiyal2024} with zero-temperature reservoirs \cite{keller2020feshbach, li2018efficient}, Feshbach-tunable nonlinearities, spin–orbit coupling \cite{li2022quantum}, or mixtures of atomic species \cite{niedenzu2019quantized}. A thermodynamic framework for nonlinear qubits, interpreting nonlinear dynamics as effective many-body behavior, was recently developed in \cite{deffner2025}. Finally, in \cite{AmetteEstrada2024a}, an Otto cycle was simulated using a finite-temperature weakly coupled description: the truncated Gross-Pitaevskii equation was used to evolve the system during adiabatic strokes, and Stochastic Ginzburg-Landau dynamics was used for isochoric thermalization. Consistent with previous studies, increased interaction strength was found to reduce efficiency, while enabling substantial increases in power.

In the present work, we implement a thermodynamic cooling cycle in the same spirit as in \cite{Gluza}, but in a fully three-dimensional and nonlinearly coupled case, in a set up composed of three spatially distinct condensates that are allowed to exchange energy and mass throughout the cycle. The protocol consists of compression and expansion strokes applied to a piston-like region, interspersed with contact operations with the system and a reservoir. The dynamics are modeled via direct numerical simulations of the truncated Gross–Pitaevskii equation, using time-dependent trapping potentials to control the strokes and condensate contacts. To characterize the thermodynamics of the set up, we extract the effective temperature of each condensate at every stage using momentum-distribution diagnostics. We find that, despite mass transfer and sound excitations, the protocol achieves successful cooling during two consecutive cycles (with the possibility of repeating the process further), with diminishing returns with every consecutive cycle.

These results suggest that cooling cycles can be studied in BECs under more realistic conditions, and open the door to further investigations involving optimal-control protocols, shortcuts to adiabaticity, and multi-cycle operation. Our work extends and refines the methodology introduced in \cite{AmetteEstrada2024a}, bringing tools traditionally used in quantum turbulence \cite{Nore1997,AmetteEstrada2022a,AmetteEstrada2022b,Shukla2019} to the study of far-from-equilibrium thermodynamic processes in BECs, while also presenting a thermometry method capable of extracting temperatures for individual condensates.

\section{Methodology}
\label{sec:mthods}

We perform numerical simulations of thermodynamic cycles using three Bose-Einstein condensates---hereafter referred to as the system, piston, and reservoir---arranged along the $z$ axis, in this spatial order, confined in an elongated three-dimensional cigar-like trap and separated from each other by time-dependent external potentials. By compressing, expanding, and performing contact operations, we aim at cooling the system, extracting heat from it and transferring it to the reservoir. Below we describe the approach used to model the set up.

\subsection{Models for condensates at finite temperature}
\label{sec:gpe}

We describe the dynamics of the three condensates using a single mean-field order parameter $\psi(\mathbf{r},t)$. To model its evolution we use the Gross-Pitaevskii equation (GPE), which describes weakly interacting Bose-Einstein condensates \cite{GPE_1,GPE_2,GPE_3}. The equation is
\begin{equation}
    i\hbar \frac{\partial \psi (\mathbf{r},t)}{\partial t}
    = \left[ -\frac{\hbar^2 \nabla^2}{2m} + g|\psi(\mathbf{r},t)|^2 + V(\mathbf{r},t) \right] \psi(\mathbf{r},t),
    \label{eq:GPE}
\end{equation}
where $m$ is the atomic mass, and $g = 4\pi \hbar^2 a / m$ is the interaction strength with $a$ the $s$-wave scattering length. In practice, this equation is truncated to a finite number of Fourier modes, and as a result what we solve is the truncated GPE, which allows us to model finite-temperature states provided initial conditions are prepared properly \cite{Proukakis2008, Berloff_2014, Shukla2019, AmetteEstrada2024a}.
The potential $V(\mathbf{r},t)$ is cylindrically confining, with rotation symmetry along the $z$ axis, and with two barriers perpendicular to $z$ that separate the condensates from one another. The explicit functional form of the trapping potential is provided in Appendix~\ref{sec:appendix B}.

The set up must first be initialized in an equilibrium state at a finite temperature. To this end  we solve the stochastic Ginzburg–Landau equation (SGLE). This equation can be obtained by applying a Wick transformation $t\rightarrow -it$ to Eq.~\eqref{eq:GPE}, adding a chemical potential $\mu$, and random forcing. This yields
\begin{multline}
         \frac{\partial \psi (\textbf{r},t)}{\partial t} = \left[ \frac{\hbar \nabla ^2}{2m} - \frac{g}{\hbar}|\psi (\textbf{r},t)|^2 - \frac{V(\textbf{r},t)}{\hbar} + \frac{\mu}{\hbar}\right] \psi (\textbf{r},t) \\
         + \sqrt{\frac{2}{\mathcal{V} \hbar \tilde{\beta}}}\zeta (\textbf{r},t).
    \label{eq:SGLE}
\end{multline}
Here $\zeta(\textbf{r},t)$ is a delta-correlated random process such that $\left< \zeta(\textbf{r},t)\zeta ^*(\textbf{r}',t')  \right>= \delta (\textbf{r}-\textbf{r}')\delta (t-t')$, and the factor $(2/\mathcal{V}\hbar \tilde{\beta})^{1/2}$ (where $\mathcal{V}$ is the total volume) sets the fluctuations' amplitude through the inverse ``temperature'' $\tilde{\beta}$. When the inverse of this coefficient is set to zero, the SGLE relaxes asymptotically to steady states of the GPE. For nonzero noise, Eq.~(\ref{eq:SGLE}) is equivalent (in Fourier space) to a Langevin equation for the evolution of each Fourier mode of $\psi (\textbf{r},t)$, and has an associated multivariate Fokker–Planck equation whose solutions converge to finite-temperature equilibrium states of the Grand-canonical ensemble \cite{Krstulovic_2011, AmetteEstrada2024a}. In our case, following the procedure described in \cite{Nore1997, AmetteEstrada2024a}, we couple the SGLE with an auxiliary equation for the mean density, to fix the total mass in the condensate (instead of the chemical potential) and thereby sampling finite-temperature solutions from the canonical ensemble.

\subsection{Numerical methods and reference values}
\label{sec:numerico}

We perform direct numerical simulations of both the SGLE and the truncated GPE (in the following, just ``GPE''), to respectively generate an initial state and to later evolve the set up in time. To this end we use a pseudo-spectral Fourier-based method in a spatial grid with $64 \times 64 \times 512$ points, employing the standard $2/3$-dealiasing rule to enforce numerical stability and conservation laws \cite{Nore1997}. The GPE is integrated using a fourth-order Runge-Kutta scheme, whereas the SGLE is advanced in time using a forward Euler method. For both equations we use the parallel code GHOST \cite{Mininni2011, Rosenberg_2020}, which is publicly available, in a box of size $[0,\pi/2]L \times [0,\pi/2]L \times [0,4\pi]L$, where $L$ denotes the unit length scale.

All quantities are reported in dimensionless units, or using units defined by a characteristic speed $U$, the unit length scale $L$, the associated time scale $\tau = L/U$, and a reference density $\rho_0$. The speed of sound is $c = \sqrt{g \rho_0 / m} = U$, and the condensate healing length in the simulations is $\xi = \hbar / \sqrt{2mg \rho_0} = 0.0707\,L$. Quantities in the following sections can be rescaled by multiplying by typical laboratory values. As an example, for typical experimental parameters, one has $L \approx 10^{-4}\ \text{m}$ and $c = U \approx 2 \times 10^{-3}\ \text{m/s}$, yielding a dimensional healing length of $\xi \approx 1.12 \times 10^{-6}\ \text{m}$. Finally, in ultracold atom experiments using, e.g., $^{87}\text{Rb}$, representative peak densities are of the order of $10^{13}\ \text{cm}^{-3}$.

\begin{figure}
    \centering
    \includegraphics[width=1\linewidth]{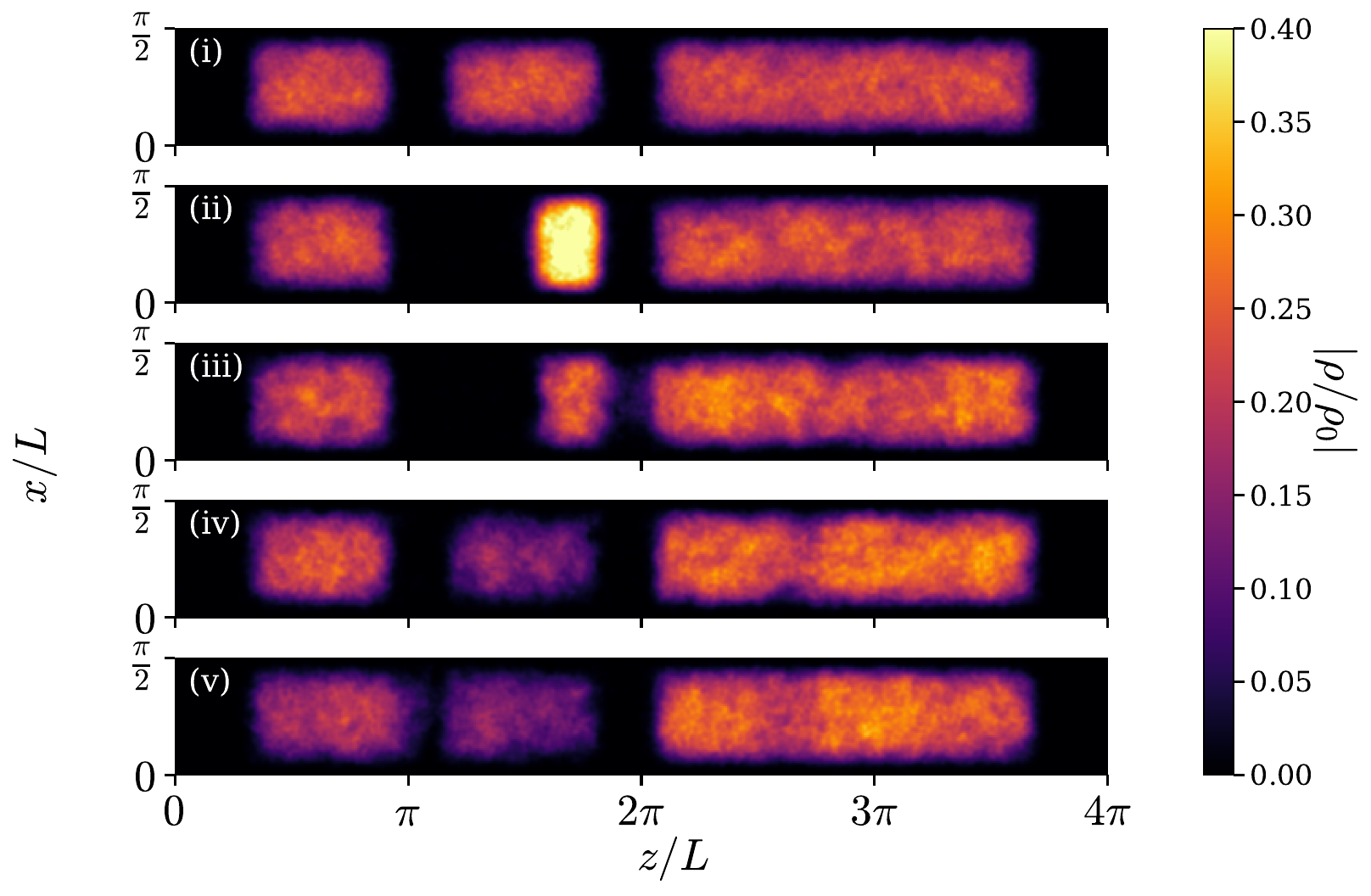}
    \caption{Representation of the thermal cycle. The panels show in color the normalized mass density $\rho /\rho _0$ in a two-dimensional slice of the condensates. From top to bottom, the figure depicts: (i) The initial state of each condensate, showing from left to right the system, the piston, and the reservoir. (ii) The compression stage of the piston. (iii) The interaction of the piston with the thermal reservoir. (iv) The subsequent expansion of the piston back to its original volume. (v) The interaction between the piston and the system, before their final decoupling. The cycle concludes after an additional integration for $20 \tau$ to reach the equilibrium.} 
    \label{fig:figure 1}
\end{figure}

\subsection{Construction of the thermodynamic cycle}
\label{sec:ciclo}

We consider a four-stroke cycle, with each stage characterized by a prescribed variation of the external potential. These variations generate nontrivial dynamical responses due to interactions, the non-equilibrium nature of the set up, and its inherent nonlinearity.

The cycle begins from an initial configuration in which the target system, the piston, and heat reservoir are all prepared at the same temperature, this is achieved by solving the SGLE at some given $\beta$. The system and piston initially have equal sizes, while the ``reservoir'' is made three times larger to mimic a thermal reservoir (see Fig.~\ref{fig:figure 1}). Afterwards, the cycle is evolved by integrating the GPE in four stages. {\it Stage one:} In the first cycle stroke, the piston undergoes an adiabatic compression, thereby increasing its energy. {\it Stage two:} In the second stroke, the compressed piston is coupled to the reservoir, by decreasing the amplitude of the potential barrier that separates them (see Table~\ref{tab:cycle_stages}). This enables heat transfer from the piston to the reservoir, driven by the induced temperature gradient. Once the established contact time has elapsed, both condensates are decoupled (by increasing again the amplitude of the potential barrier that separates them to its original value). {\it Stage three:} The piston is then adiabatically expanded in a controlled manner back to its initial volume, reducing its energy. {\it Stage four:} In the fourth stroke, the piston is coupled to the target system (using the same methodology as in stage two), allowing an energy exchange between them. Finally, these two condensates are decoupled, and the set up is allowed to evolve freely, undergoing relaxation toward equilibrium. Note that $V({\bf r}, t)$ changes smoothly in time on all stages. The parameters used for each stage are listed in Table~\ref{tab:cycle_stages}.

\begin{table}
\centering
\begin{tabular}{c c c c c c}
\toprule
\textbf{Stage} & \textbf{Operation} & \textbf{Duration} & \textbf{Piston length} & \textbf{Barrier} \\
\midrule
1   & Compression    & $20\,\tau$  & 100$\%$-60$\%$     & 100$\%$ \\
1   & PR contact     & $180\,\tau$ & 60$\%$             & 30$\%$  \\
3   & Expansion      & $20\,\tau$  & 60$\%$-100$\%$     & 100$\%$ \\
4   & PS contact     & $180\,\tau$ & 100$\%$            & 10$\%$  \\
    & Relaxation     & $20\,\tau$  & 100$\%$            & 100$\%$ \\
\bottomrule
\end{tabular}
\caption{Stages of the thermodynamic cycle, with the operation performed in each stage, where for the contacts ``PR'' stands for Piston-Reservoir and ``PS'' stands for Piston-System. Changes in the length of the piston (in the $z$ direction) are indicated with percentages, e.g., 100$\%$-60$\%$ means the piston was compressed from its original length to 60$\%$ of its length. The column ``Barrier'' gives the amplitude of the barrier in percentage of its original value, e.g., 30$\%$ means the amplitude is decreased to 30$\%$ of its initial value.}
\label{tab:cycle_stages}
\end{table}

\subsection{Thermometry of each condensate}

In order to analyze the cooling cycle, we need a method to quantify temperature in each individual condensate as a function of time. Although temperature estimation is a well-studied problem in experimental setups \cite{Gerbier, Davis}, such measurements typically rely on destructive methods as, e.g., releasing the cloud and reconstructing the momentum distribution. In numerical simulations, however, temperature is usually inferred only at a global level, often through mode occupation counting, or measuring the mean central density \cite{Shukla2019, AmetteEstrada2024a, AmetteEstrada2025a}. Here we need to measure a temperature for each condensate. To this end we use the momentum distribution. In GPE, the momentum density (per unit volume) can be obtained from the order parameter as \cite{Nore1997, Shukla2019}
\begin{equation}
    \mathbf{p} = -\frac{i\hbar}{2}\left( \psi^* \boldsymbol{\nabla} \psi - \psi \boldsymbol{\nabla} \psi^* \right),
    \label{eq:momentum}
\end{equation}
From Eq.~(\ref{eq:momentum}) we compute instantaneous probability distribution functions (PDFs) of momentum in each condensate.

\begin{figure}
    \centering    
    \includegraphics[width=0.99\linewidth]{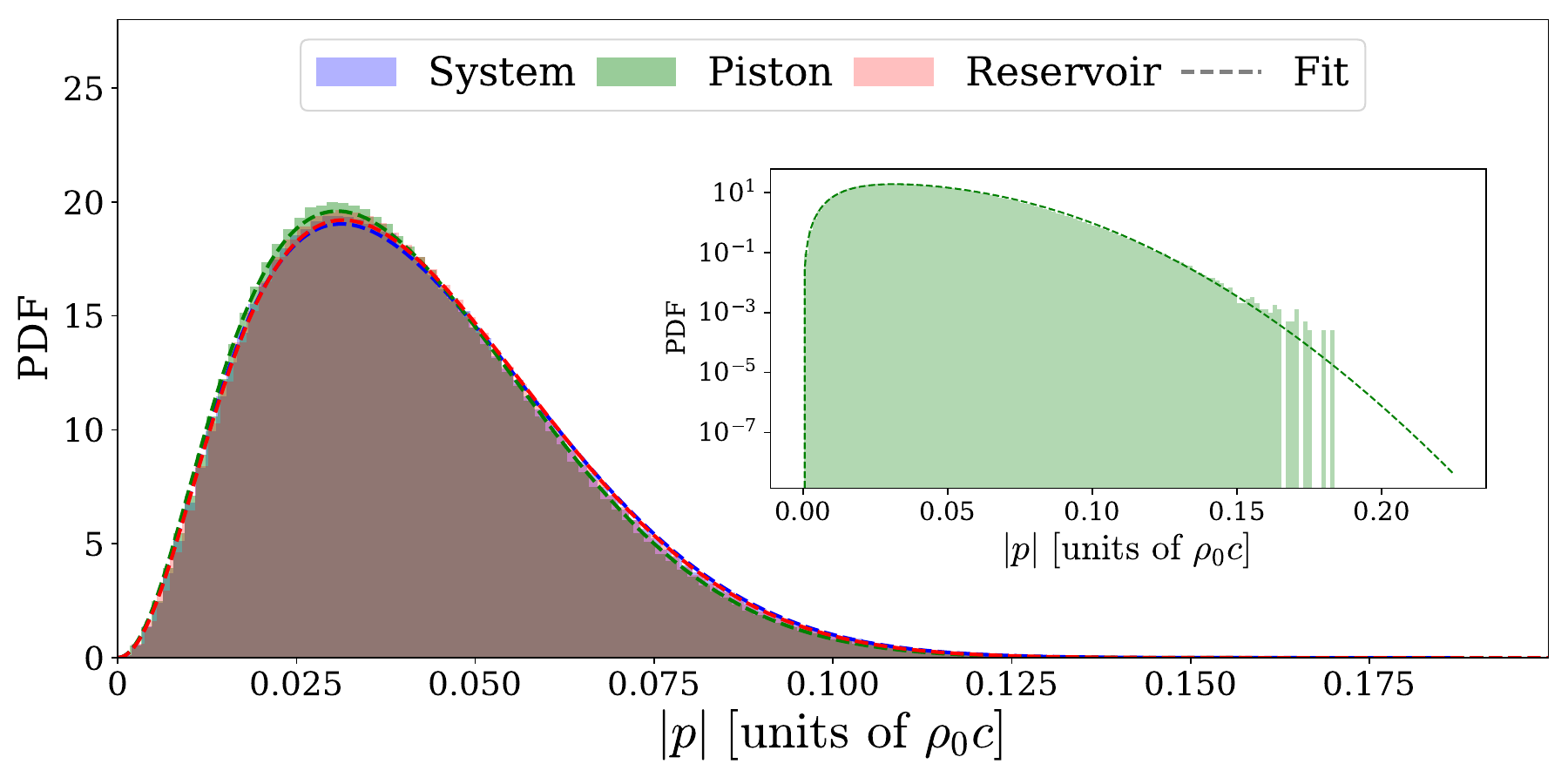}
    \caption{Initial PDFs of momentum for the three condensates with their respective fits using Eq.~(\ref{eq:momentum BE}), showing good agreement between the theoretical expression and the equilibrium obtained through the preparation of the initial thermal bath. Note the three condensates (system, piston, and reservoir) have similar PDFs. The inset shows the PDF of the piston in semilog scale, to showcase the agreement also in the tail of the PDF. Results for the system and the reservoir are analogous.}
    \label{fig:figure 2}
\end{figure}

In the semiclassical description of a non-interacting Bose gas, the momentum distribution is in general \cite{Pat}
\begin{equation}
f(p) = \frac{p^2}{e^{\beta (p^2 + \alpha)} - 1},
\label{eq:momentum BE}
\end{equation}
where $p = |\bf {p}|$, $\beta$ is an inverse temperature, and $\alpha$ includes the contribution of the chemical and spatial potentials. Although the expression is derived for a non-interacting condensate, it still applies to our system as we see in Fig.~\ref{fig:figure 2}, probably because our set up, albeit interacting, is sufficiently diluted, and the trapping potential is zero everywhere except near the borders of the cylindrical trap and the barriers (i.e., the gas in each condensate is more or less homogeneously distributed, as seen in Fig.~\ref{fig:figure 1}).

Under these assumptions we can then estimate $\beta$ or $T$ from a best fit of Eq.~(\ref{eq:momentum BE}) to $f(p)$ obtained from the GPE simulations for each condensate. In practice, we will provide $T/T_c$ or $\beta_c/\beta$, i.e., the temperature (or the inverse temperature) in units of the critical temperature (or inverse critical temperature) of the condensate. The critical values were estimated in two different ways: On the one hand, we performed several numerical simulations of the SGLE with different noise amplitudes $(2/\mathcal{V} \hbar \tilde{\beta})^{1/2} \, \zeta({\bf r},t)$, and used the method of counting the occupation of low-momentum levels as the temperature was varied to determine $\beta_c$ \cite{Shukla2019}, or equivalently, of measuring the mean central density of the condensate as the temperature was varied \cite{AmetteEstrada2025a}. On the other hand, we implemented a new method based on comparing the prediction for $1/\beta$ from Eq.~(\ref{eq:momentum BE}) with the value used for the noise amplitude in SGLE. These two values should be linearly dependent as long as the set up is subcritical \cite{Proukakis2008}, allowing a second determination of $\beta_c$. All methods yield similar estimations for the critical temperature, and details on the second method are provided in Appendix~\ref{sec:appendix A}.

\section{Cycle evolution and cooling}
\label{sec:results}

We describe first the detailed evolution of each condensate during each individual stroke of a single cycle, with emphasis on the temperature changes in each component of the set up. Aftwerwards, we consider a second consecutive cycle and its effects. 

\subsection{Initial state preparation and compression}

\begin{figure}
    \centering
    \includegraphics[width=1\linewidth]{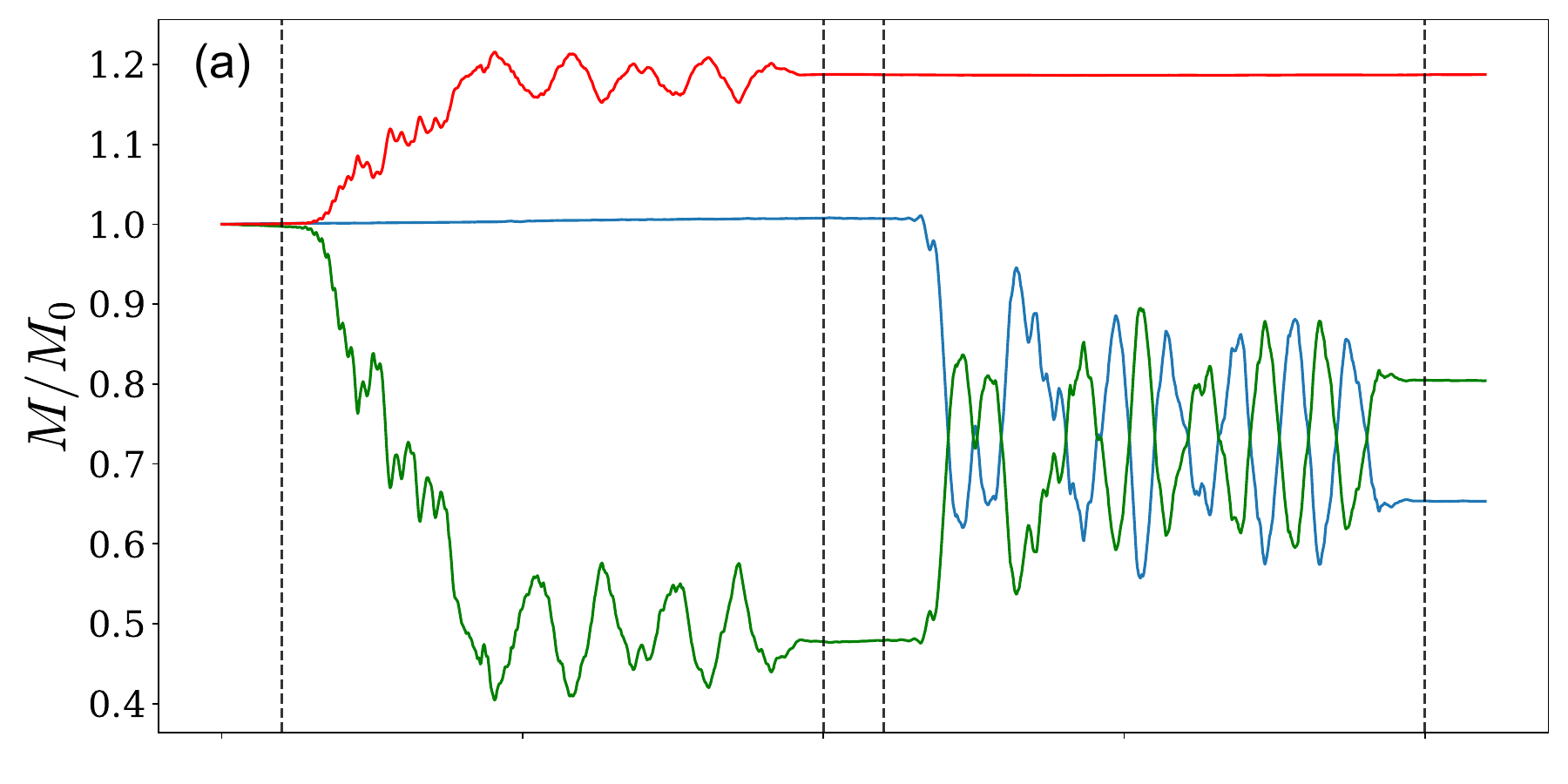}\\
    \includegraphics[width=1\linewidth]{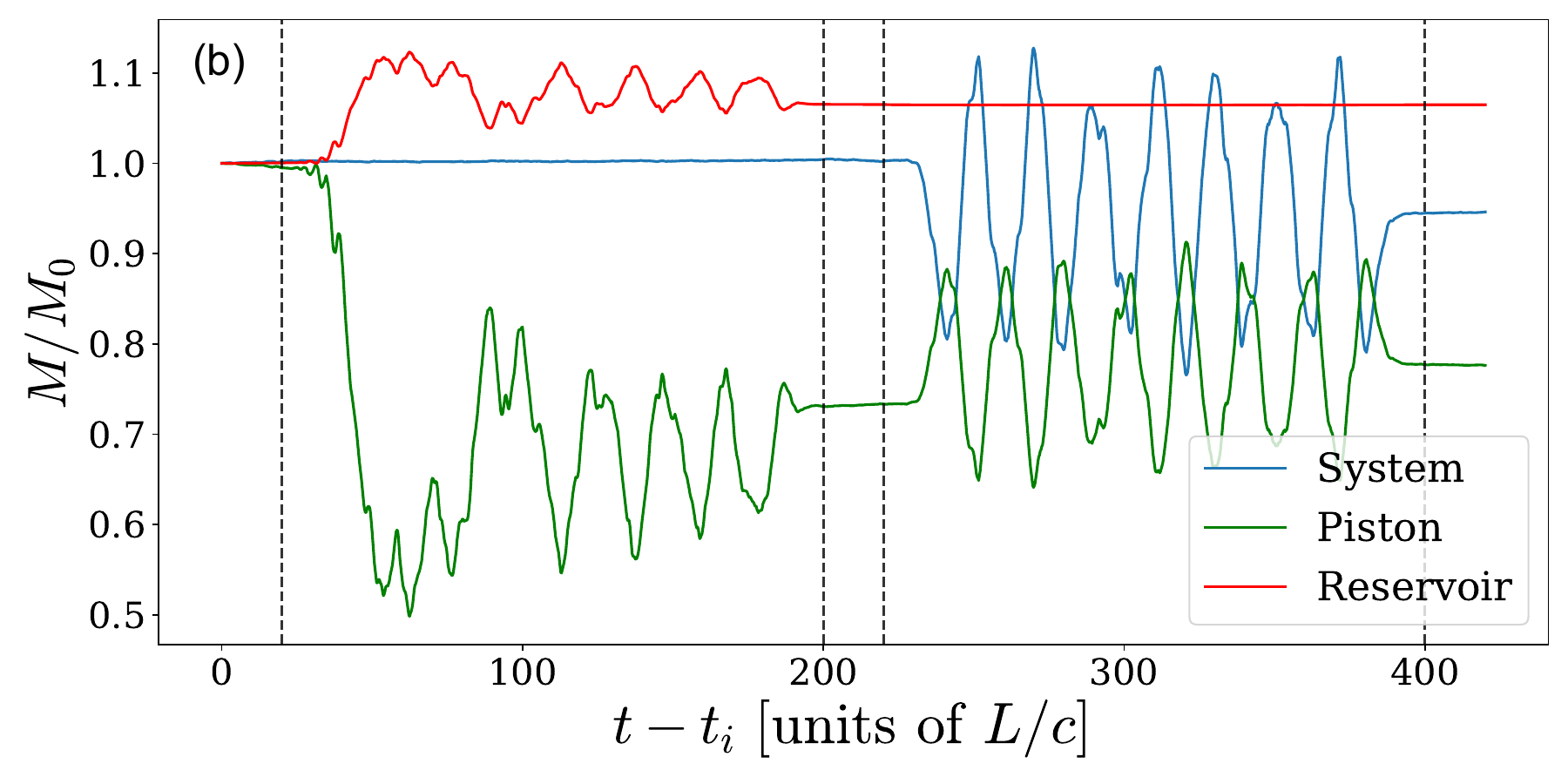}
    \caption{Mass dynamics of each condensate during the first (a) and second (b) cycle. 
    Each panel shows the evolution of the mass in each condensate with respect to its initial mass, $M/M_0$ (where $M_0$ is the mass at the beginning of each cycle); $t_i$ is the initial time of each cycle. The dashed lines separate the (1) compression, (2) PR contact, (3) expansion, and (4) PS contact, followed by a final relaxation (see Table~\ref{tab:cycle_stages}). The piston undergoes substantial mass transfer followed by oscillations during each contact stage, specially on the first cycle.}
    \label{fig:figure 4}
\end{figure}

We consider an initial thermal state with the three condensates at $T_0=0.029\ T_c$ (we verified that using other initial temperatures for the three condensates gives similar results). Figure \ref{fig:figure 2} shows $f(p)$ for the three condensates in the set up at $t=0$, and the fit from Eq.~(\ref{eq:momentum BE}). The PDFs of the three condensates are very similar, differing only by small fluctuations arising from the generating process. Owing to the properties of the set up, these fluctuations relax rapidly after a short evolution of the GPE.

\begin{figure*}
    \centering
    \includegraphics[width=.9\linewidth]{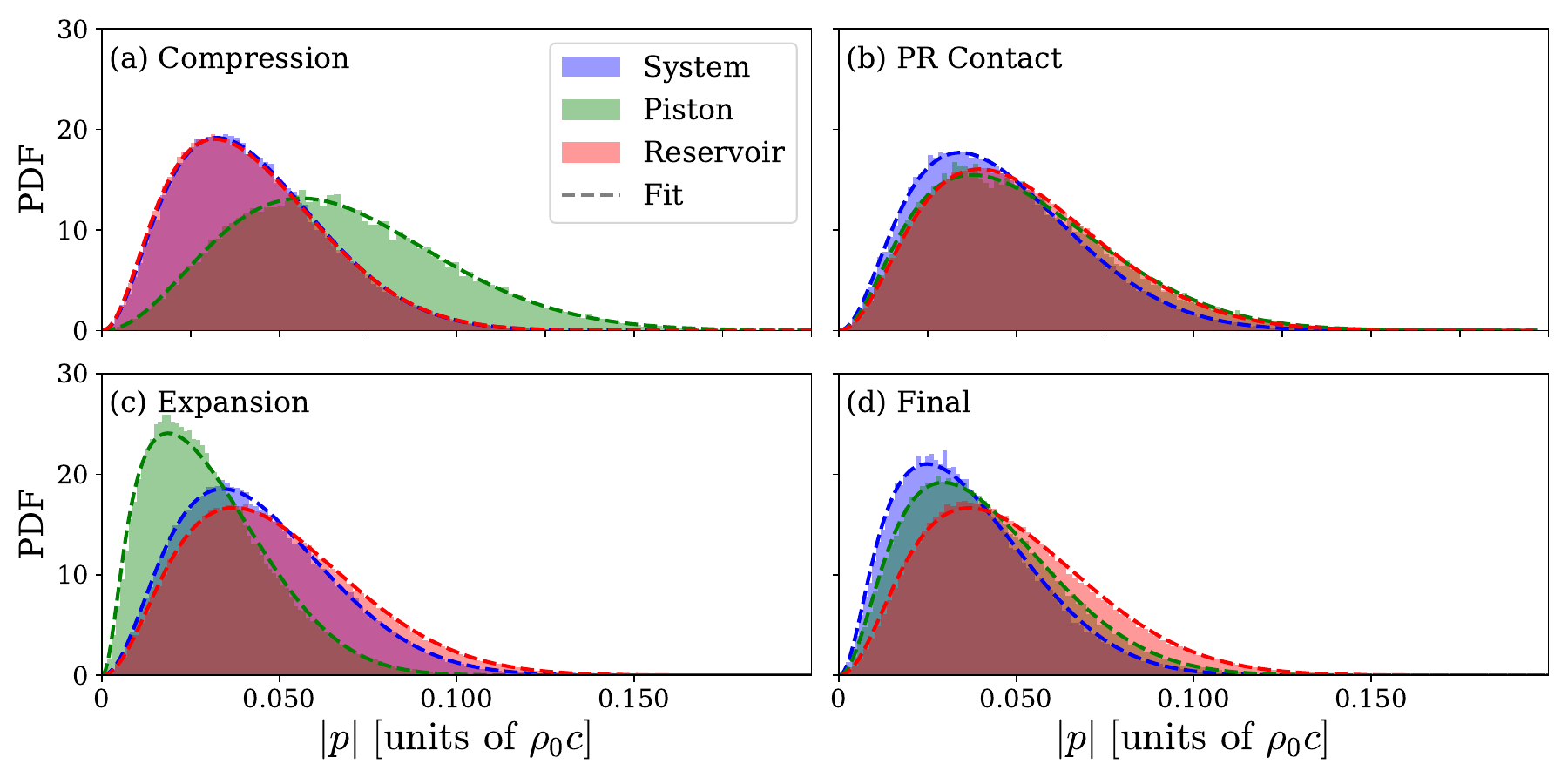}
    \caption{PDFs of momentum with their corresponding best fits using Eq.~(\ref{eq:momentum BE}) for the system (blue), piston (green), and reservoir (orange). From top left to bottom right the PDFs correspond to the end of the (a) compression stage, (b) piston-reservoir contact stage, (c) expansion stage, and (d) final state after relaxation.}
    \label{fig:figure 3}
\end{figure*}

From this initial condition, the piston is compressed adiabatically to $60\%$ of its initial length (and volume) over $20\,\tau$. This is done by widening the potential barrier that separates the piston from the thermal reservoir, thereby performing work on the former while maintaining the system and reservoir isolated. The choice of parameters (in particular, the compression speed) is such that the piston evolution is quasi-adiabatic. Under this condition, changing the compression speed does not affect the final state (we verified this explicitly). The mass exchange between the piston and the reservoir throughout this stage is negligible, as can be seen in Fig.~\ref{fig:figure 4}(a). This suggests that our condensates are well separated by the barriers, and that each condensate can be studied individually.

The PDFs of momentum of each condensate at the end of the compression, with their best fits using Eq.~(\ref{eq:momentum BE}), are shown in Fig.~\ref{fig:figure 3}(a). The broadening of the piston’s momentum distribution, along with the shift in the position of its maximum, indicates that the gas is heated during the compression process. In particular, the piston reaches a temperature $T_\text{comp} \approx 0.062\ T_c$, i.e., $T_{\text{comp}} / T_0 \approx 2.14$ higher than its initial state. In contrast, both the system and the reservoir PDFs remain unchanged.

\subsection{Piston-Reservoir Contact}
\label{sec:PR}

This stage couples the compressed and heated piston with the reservoir, resulting in heat transfer from the former to the latter. This is accomplished by reducing the amplitude of the potential barrier that separates them, thereby increasing its permeability. Under ideal conditions, heat transfer should proceed gradually with negligible mass transfer, provided that the interaction between the two regions remains sufficiently weak. This, in turn, requires a very small contact area. The caveat of this is that weak contact requires very long times to reach thermalization between the two condensates, while a stronger contact can lead to faster thermalization (and thus, increased power in the entire cycle) at the cost of larger mass exchange. The selection of the contact conditions thus define strongly the cycle efficiency.

The contact protocol used consists in a harmonic lowering and upbringing of the potential barrier, keeping the barrier static on its lowest amplitude for a long period of time before bringing it back up, such that both condensates can relax. The barrier amplitude is first modulated by $A(t) = 1- \Delta \sin ((t-t_{c1})/t_{ud})$ for the downward motion, where $\Delta$ controls the decrease in the amplitude, $t_{c1}$ is the time at which the contact starts, and $t_{ud}$ is the time it takes for the barrier to go up or down. Afterwards, the barrier remains low and static for a time $t_s$, and then it is brought back up with amplitude $A(t) = 1- \Delta \cos ((t-t_{c2})/t_{ud})$, where $t_{c2} = t_{c1} + t_s$. Note the barrier evolution is reversible in time. We use $180\,\tau$ as the total time of this stroke, with $t_{ud}=20\,\tau$ and $t_s=140\,\tau$. As already mentioned, the barrier is lowered to $30\%$ of its initial amplitude. That choice was based on a systematic sweep of this parameter, ranging from $80\%$ to $20\%$ of the barrier's initial amplitude, and we chose the amplitude that resulted in a fast equilibration of the two condensates with a reasonable mass exchange.

The mass evolution in each condensate is shown in Fig.~\ref{fig:figure 4}. There, we can observe an increase in the reservoir mass and a decrease in the piston mass during the second stage of the first cycle (relative to their respective initial conditions), followed by counter-phase oscillations between the piston and reservoir, which reveal the complex dynamics of the contact. These oscillations, corresponding to sound waves excited during the contact and traveling through the condensates, are modulated by the barrier evolution, growing when the barrier is lowered and decreasing when it is raised. At the end of the contact the piston loses approximately $55\%$ of its mass, which is transferred to the reservoir. Thermalization between the piston and the reservoir can be confirmed in Fig.~\ref{fig:figure 3}(b), as both histograms overlap with a temperature larger than the initial condition of the reservoir but lower than that of the compressed piston. Both condensates reach $T_\text{contact} \approx 0.046 \ T_c$, or $T_\text{contact}/T_0 \approx 1.64$.

\subsection{Expansion}
\label{sec:expansion}

At the beginning of this stage we have, again, three condensates separated by potential barriers.
We proceed to reverse the process done in the first stage, expanding the piston back to its original volume over a time span of $20 \ \tau$. Again, this process only affects the piston, while both the reservoir and system remain undisturbed. 
As a result of the expansion, see Fig.~\ref{fig:figure 3}(c), the piston ends up cooler at a temperature $T_\text{expansion}/T_0 \approx 0.65$ colder than its initial state. As the piston ends in a state markedly colder than the system, we can expect that putting them in contact should result in system cooling.

\subsection{Piston-System Contact and relaxation}
\label{sec:PS}

In these stages the system and piston are set in contact with each other to thermalize, and finally a relaxation of the entire set up with the barriers up is allowed. For the contact stage, parameters are changed from those used in the piston-reservoir contact, as the mass loss of the piston in that stage results in lower interaction (or thermal conductivity). In other words, we need a more permeable barrier for the piston.  Therefore, the barrier amplitude between the piston and the system is lowered to 90$\%$ of its initial value.

Figure~\ref{fig:figure 4}(a) depicts a similar behavior of the mass evolution as in the previous contact phase. Oscillations, this time larger and between the piston and the system, appear when both condensates make contact after a quick mass exchange. The oscillations again reflect the presence of mass density waves that move along the condensates back and forth. They are enhanced by the larger mass imbalance between these two condensates. During this process, the piston's density grows, ending with approximately 80$\%$ of its initial mass, while the system loses about 30$\%$ of its mass. This is to be expected as the imbalance in both densities generates a big mismatch in the chemical potential of both condensates.

\begin{figure}
    \centering
    \vspace{0.5cm}
    \includegraphics[width=1\linewidth]{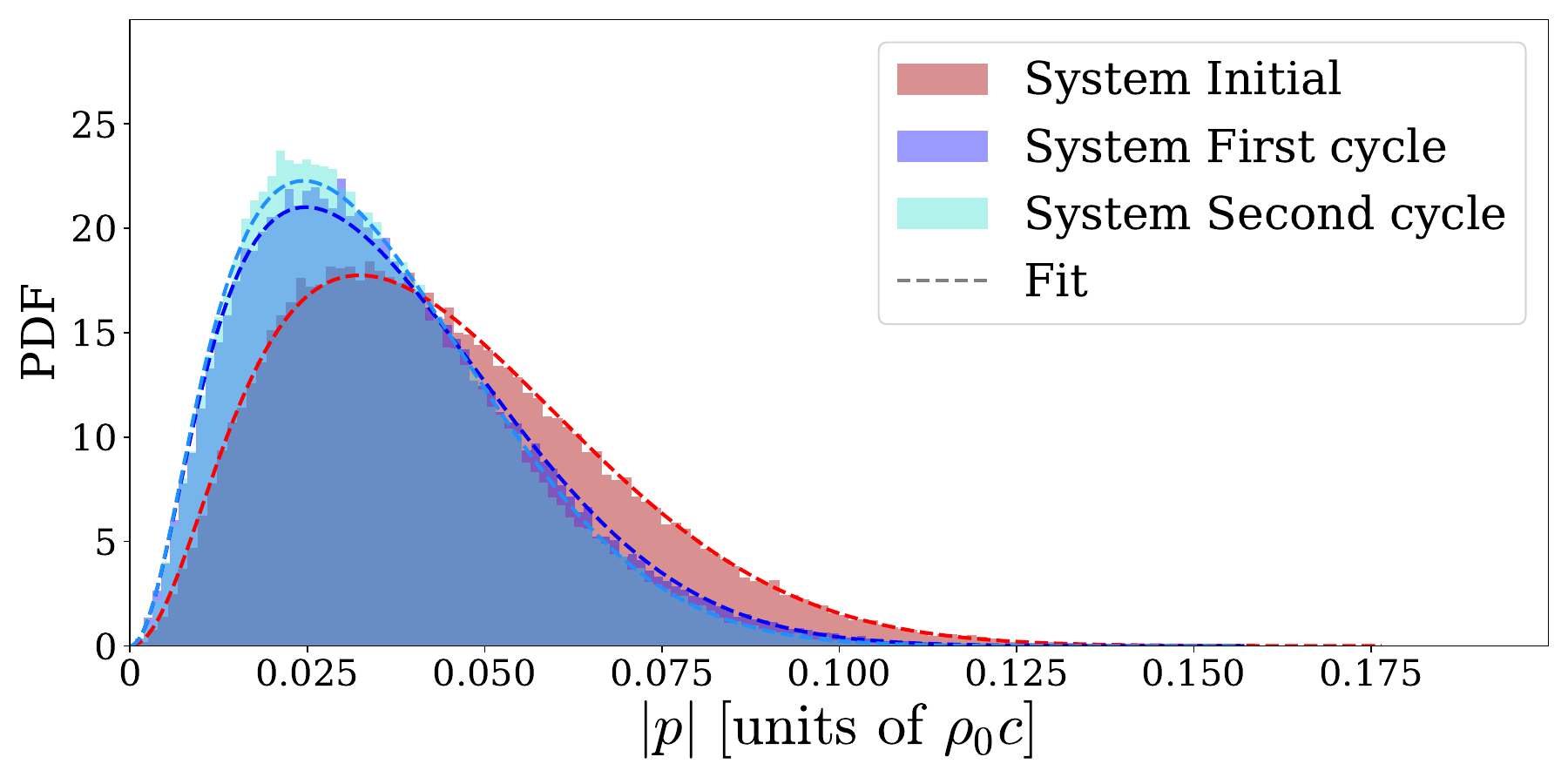}
    \caption{PDFs of momentum for the system at the beginning ($t =0$) of the first cycle (red), at its end ($t = 420 \ \tau$, in blue), and at the end of the second cycle ($t=840\ \tau$, in cyan). The dashed lines indicate the best fit using Eq.~(\ref{eq:momentum BE}).}
    \label{fig:figure 5}
\end{figure}

The PDF of momentum for each condensate after the contact ends, and after the final relaxation, is shown in Fig.~\ref{fig:figure 3}(d). 
The system cools down and ends even with a slightly smaller temperature than the piston. This behavior is associated to the cycle protocol and the sound waves generated during contact. Raising the barrier slightly earlier or later can induce small temperature differences on both sides. In principle, adjusting the timing of these oscillations could be employed to further cool down the system. A detailed comparison between the system's initial and final PDFs of momentum is shown in Fig.~\ref{fig:figure 5}. Note that the system underwent an effective cooling process during the whole cycle with $T_\textrm{final} / T_0 \approx 0.83$.

It is worth highlighting that the cycle is done in finite time, allowing for the development of three-dimensional structures in the condensates, with waves and density fronts that propagate along them exciting complex dynamics. As a result, compressions, expansions, and contact processes in the cycle are imperfect, inducing phenomena which have not been considered in previous works. In spite of this, the final effect is still the desired for a thermal refrigerator. The protocol seems possible to reproduce in experimental setups, where assumptions such as the absence of mass transfer or quasi-stationary can be unrealistic. Moreover, our approach allows us to consider possible ways to obtain cooling enhancing by, e.g., adiabatic shortcuts that exploit synchronization of the cycle with the system internal dynamics.

\subsection{Second cycle}
\label{sec:Cycle 2}

It is natural now to ask whether this procedure can be repeated to achieve further cooling. The iteration of the process is limited by the particular parameters of the cycle, and may be arrested by mass imbalances, e.g. cases in which the system or the piston have excessive mass losses, rendering the engine useless for practical applications. But in spite of these limitations, we still observe a successful second iteration of the cycle for the choice of parameters discussed here (a third iteration also results in cooling, albeit with a diminishing return). For the second iteration, the main difference with respect to the first cycle is shown in Fig.~\ref{fig:figure 4}(b). The system losses only $10\%$ of its initial (second cycle) mass, instead of $30\%$ in the first cycle (this implies a compounded loss of around $63\%$ of the initial mass over the two cycles).

In Fig.~\ref{fig:figure 5}, we show the momentum PDFs of the system at the initial state ($t=0$), at the end of first cycle, and at the end of the second cycle, where $T^{(2)} _f \approx 0.022\ T_c$. In other words, $T^{(2)} _f / T^{(1)} _f \approx 0.88$ ($T^{(1)} _f$ is the final temperature at the end of the first cycle). This results in a total effective cooling of $T^{(2)} _f/T_0 \approx 0.73$ after two cycles, showing that the set up loses cooling power throughout the second cycle, as expected, but can still cool the system down.

\section{Conclusions}
\label{sec:conclusions}

In this work we showed that a fully three-dimensional, weakly interacting Bose–Einstein condensate can sustain a complete thermodynamic cooling cycle, driven solely by controlled modulations of a trapping potential. By combining SGLE-based finite-temperature initialization of thermal baths, with numerical simulations of the GPE, we were able to capture the nonlinear, non-equilibrium dynamics of a quantum thermal machine in the mean field, semi-classical approximation, operating across spatially separated condensates. This includes performing operations on the condensate in finite time, the observation of density waves being excited by the external potentials, and their subsequent propagation and thermalization.

During the first cycle, we are able to cool the target system about $20\%$ from its initial temperature, where the temperature was determined from momentum PDFs. We further showed that the protocol is capable of performing at least one more cycle, with the second iteration still providing additional, yet less effective, cooling of the system. This shows that the procedure can be repeated and is not restricted to a single pass, although diminishing returns emerge naturally, as expected. In particular, the second cycle lost $\approx 5\%$ cooling effectiveness compared to the first one, but the overall result after two cycles resulted in an accumulated effective cooling of $\approx 27\%$ with respect to the starting state. It is worth mentioning that, although we present here a cycle with a particular choice of parameters, we did a broad exploration of parameter space. The cycle works for different choices of the initial temperatures of the set up. We also explored different choices for the duration of each stroke, and for the parameters that characterize each stroke (e.g., the piston compression, or the change in the amplitudes of the barriers). In this way, the methodology presented here can be used to design other cycles in the future. 

More broadly, the results highlight that mass transport along the potential barriers is not an undesirable artifact of these kind of cycles, but can be used as a functional component of the cooling mechanism. The cycle relies on the spatial organization of excitations, the interplay between compressibility and hydrodynamic flow, and the geometry imposed by the trap. As such, the phenomena observed here belong to a genuinely many-body, finite-time, three-dimensional regime that departs substantially from simplified effective models. This also opens the door for the development of adiabatic shortcuts by timing the perturbations in the system and the use of nontrivial similar protocols in experiments.

Overall, this work establishes a dynamically consistent route toward implementing quantum thermal machines in ultracold atomic gases. Beyond depicting the possibility of cooling using a particular set up, it offers a platform to explore optimal control of contact protocols, shortcuts to adiabaticity, persistent thermal cycles, and alternative refrigeration strategies in this and other geometries.

\bibliographystyle{apsrev4-2}
\bibliography{references}

\appendix
\section{Trapping potential}
\label{sec:appendix B}

The trapping potential is composed of three contributions, $V(\textbf{r}) = V_\textrm{B1} + V_\textrm{B2} + V_\textrm{trap}$. Two of them, $V_\textrm{B1}$ and $V_\textrm{B2}$, parameterize the tunable barriers that act as valves separating the condensates. The third term, $V_\textrm{trap}$, provides the overall cylindrical (cigar-shaped) confinement. The expressions for each contribution are
\begin{align}
     {}& V_{\text{B1}}(z) = \frac{V_0 A(t)}{\left[1+e^{\epsilon (z-z_1- w(t))}\right] \left[ 1+e^{- \epsilon (z-z_1+w_0)}\right]}, \\
     {}& V_{\text{B2}}(z) = \frac{V_0 A(t)}{\left[1+e^{\epsilon (z-z_2- w_0)}\right] \left[ 1+e^{- \epsilon (z-z_2+w_0)}\right]}, \\
     {}& V_{\text{trap}}(x,y,z) = 2 V_0 \left[
     \frac{1}{1+e^{-\gamma ((x^2 + y^2)^{1/2} - r_0)} } + \right. \nonumber \\
     {}& \qquad \qquad \qquad \qquad + \left. \frac{1}{1+e^{-\gamma (|z-2 \pi L| -z_0)} } \right].
\end{align}
Here, $V_0$ is an amplitude, $\epsilon$ and $\gamma$ control how fast the potentials grow, $z_1$ and $z_2$ control the central position of the barriers, $w_0$ controls the width of the barriers, $w(t)$ allows the first barrier to change width in time (to compress or expand the piston), $r_0$ and $z_0$ control the position of the borders of the cigar trap, and $A(t)$ is the parameterization of the barrier amplitude used to put the condensates in contact.

\begin{figure}
    \centering
    \vspace{0.5cm}
    \includegraphics[width=0.99\linewidth]{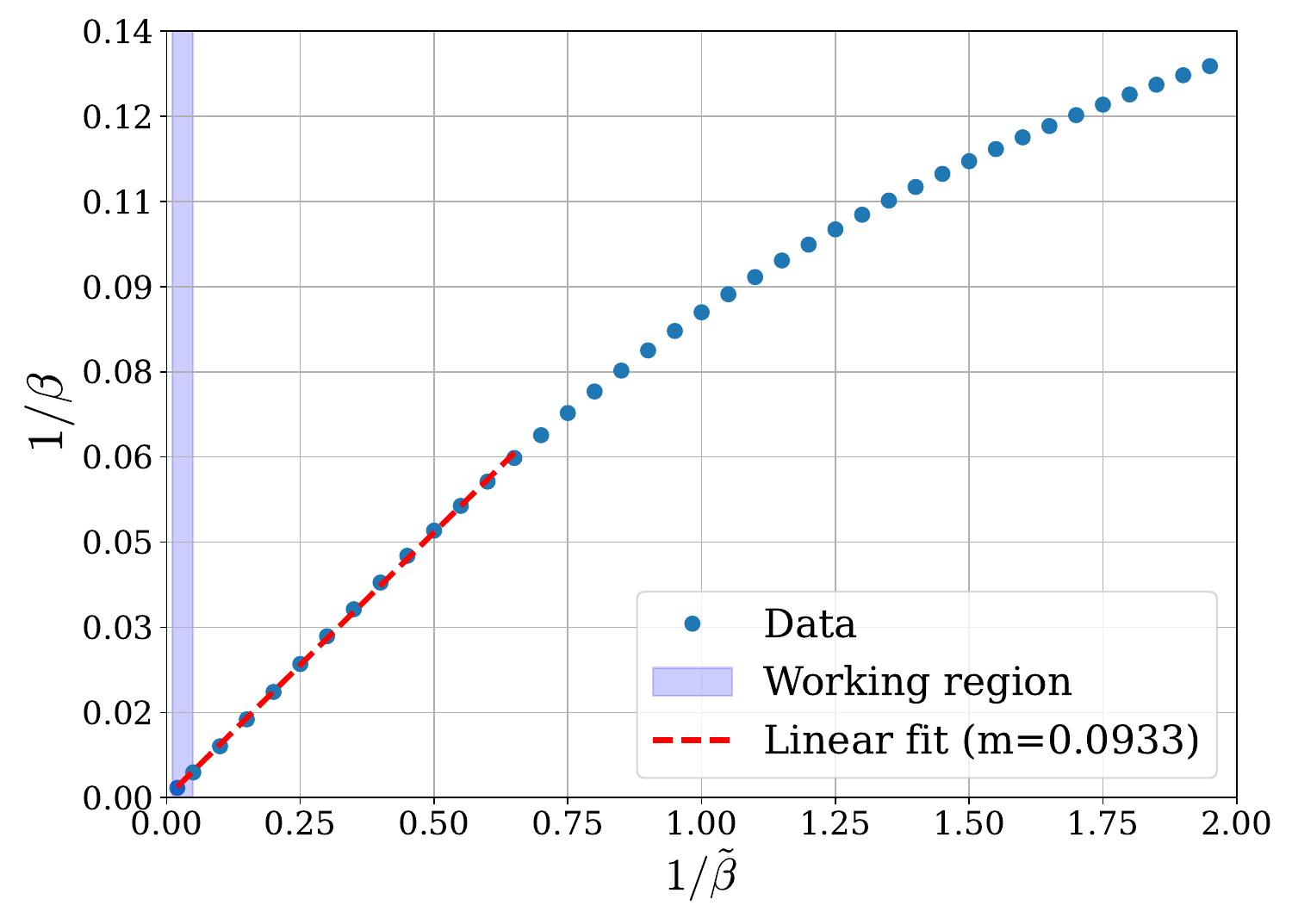}
    \caption{Calibration of the SGLE noise against the temperature inferred from the PDF of momentum. Blue points show the measured relationship between the SGLE noise amplitude $1/\tilde{\beta}$, and the temperature $1/\beta$ obtained from a best fit to the PDF (both in arbitrary units). The linear regime with $1/\beta \approx m /\tilde{\beta}$ is indicated by the red dashed fit. The shaded region marks the temperatures explored for our study.}
    \label{fig:1_beta vs T}
\end{figure}

\section{Temperature estimation}
\label{sec:appendix A}

All temperatures in this paper are expressed in terms of the condensate critical temperature. To estimate this value, we did a temperature sweep varying the noise amplitude in the SGLE with all potentials fixed in time. We recall the reader that we denote $1/\hat{\beta}$ as the ``temperature'' that controls the noise amplitude in Eq.~(\ref{eq:SGLE}). At each temperature, we measured the occupation of Fourier modes with low momentum (i.e., with low wave number $k$ in Fourier space) as done in \cite{Shukla2019}. We also measured the mean density in the center of the condensates as done in \cite{AmetteEstrada2025a}. We then obtained the critical temperature from the case in which these quantities approached zero. Then, using the same temperature sweep, we compared $1/\hat{\beta}$ in Eq.~\eqref{eq:SGLE} with the value of $1/\beta$ obtained from a best fit using Eq.~\eqref{eq:momentum BE} to the PDF of momentum from the simulations. Even though both values can have different normalizations, as long as a condensate is present they should be linearly dependent. Figure \ref{fig:1_beta vs T} shows the SGLE noise amplitude against temperature obtained from the PDF of momentum (both in arbitrary units). The values at which the data departs from the linear relation coincides with our two independent estimations of the critical temperature. From this value, and from the proportionality constant in Fig.~\ref{fig:1_beta vs T}, we can compute initial temperatures from the parameters in SGLE, and from the PDFs of momentum of each condensate, in units of the critical temperature $T_c$. Figure \ref{fig:1_beta vs T} also shows a shaded region indicating the range of temperatures explored in this work.

\end{document}